\documentclass[aps, reprint, prl, superscriptaddress]{revtex4-1}
\usepackage[caption=false]{subfig}
\usepackage{color}
\usepackage{graphicx}
\usepackage{natbib}
\usepackage{amsmath}
\usepackage{ulem}
\usepackage{braket}
\usepackage[utf8]{inputenc}
\usepackage{url}
\usepackage{hyperref}
\usepackage{amssymb}
\usepackage{bbm}

\begin{document}

\newcommand{\comment}[1]{\textcolor{blue}{#1}}
\newcommand{\commentImp}[1]{\textcolor{red}{#1}}
\newcommand{\todo}[1]{\textcolor{blue}{ToDo: {#1}}}
\newcommand{\h}[1]{\hat{#1}}
\newcommand{\sinc}{\mathrm{sinc}}

\title{N00N states from a single non-linear directional coupler}

\author{Regina Kruse,$^1$ Linda Sansoni,$^1$ Sebastian Brauner,$^1$ Raimund Ricken,$^1$ Craig S. Hamilton,$^2$ Igor Jex,$^2$ and Christine Silberhorn}
\affiliation{Integrated Quantum Optics Group, Applied Physics, University of Paderborn, Warburger Straße 100, 33098 Paderborn, Germany\\
$^2$FNSPE, Czech Technical University in Prague, B\v{r}ehov\'a 7, 115 19, Praha 1, Czech Republic}

\begin{abstract}
In the quest for applicable quantum information technology miniaturised, compact and scalable sources are of paramount importance.
Here, we present the concept for the generation of 2-photon N00N states without further post-processing in a single non-linear optical element. Based upon a periodically poled waveguide coupler, we present the principle of state generation via type-0 parametric down-conversion inside this type of devices. With the eigenmode description of the linear optical element, we utilise the delocalised photon pair generation to generate a N00N state in the measurement basis. We show, that we are able to eliminate the need for narrow-band spectral filtering, as well as for phase-stabilisation of the pump light, making this approach an elegant way to produce 2-photon N00N states.
\end{abstract}

\maketitle

\section{Introduction}

During the last decades integrated optics has become a working horse for the photonic industry. Waveguide based lasers, combined with passive beam splitters as well as active modulators have enabled a lot of progress in the optics research field.
In a more recent development, the integrated quantum optics community has strived to profit from the achievements in classical integrated optics for the miniaturisation of quantum circuits \cite{tanzilli_highly_2001, politi_silica--silicon_2008, peruzzo_quantum_2010, owens_two-photon_2011,crespi_integrated_2011, krapick_efficient_2013, corrielli_rotated_2014, metcalf_quantum_2014}, e.g. for compact quantum communication or quantum computation devices. 
However, it still remains a challenge to combine the generation of single photons with linear circuits on one chip. Recently integrated devices have been fabricated, combining the state generation with linear manipulation on a chip \cite{silverstone_-chip_2014, jin_-chip_2014}. 
This approach eliminates the incoupling losses of single photons to the linear networks, which is a prerequisite for the combination of multiple sources, e.g. for boson sampling \cite{crespi_integrated_2013, spring_boson_2013, broome_photonic_2013, tillmann_experimental_2013}. Still, the reliable preparation of sophisticated quantum states from multichannel devices requires generally a well defined and stabilised phase for the pump light. Furthermore, achieving indistinguishability of photons generated by different sources remains difficult and is usually realised by narrow-band filtering, introducing high loss in the experiment.
The integration of state generation into the linear element additionally offers many new possibilities \cite{mista_nonclassical_1997, solntsev_spontaneous_2012, kruse_spatio-spectral_2013, solntsev_generation_2014}. Firstly, it pushes the miniaturisation even further and secondly, the more important aspect, we gain access to new dynamics \cite{hamilton_driven_2014}, which are not available in conventional single photon sources. 
Recently, Lugani et al. \cite{lugani_generation_2011} proposed a scheme for the integrated generation of 2-photon N00N states. However, due to its use of propagation constant matching, it requires careful parameter design and has low tolerance for fabrication imperfections.

In this paper, we integrate photon-pair generation into a directional coupler and generate genuine 2-photon N00N states \cite{dowling_quantum_2008, afek_high-noon_2010, israel_experimental_2012, boyd_quantum_2012} without postprocessing. We eliminate the need for narrow-band spectral filtering to prepare indistinguishable photons and are able to fully forego phase-stabilised pumping of the process. By using the eigenmode description of the linear waveguide coupler, show that the photon pairs are generated into linear superpositions of the waveguide modes and that with a suitable choice of the pump frequency we are able to obtain a genuine N00N state at the output. First, we give a detailed description of the integrated system and discuss the linear properties of this device. Then, we analytically describe the concept of the state generation in this source type and investigate numerically the expected fidelity for realistic fabrication parameters.

\section{Analytical Description}

\begin{figure}
\includegraphics[width=.4\textwidth]{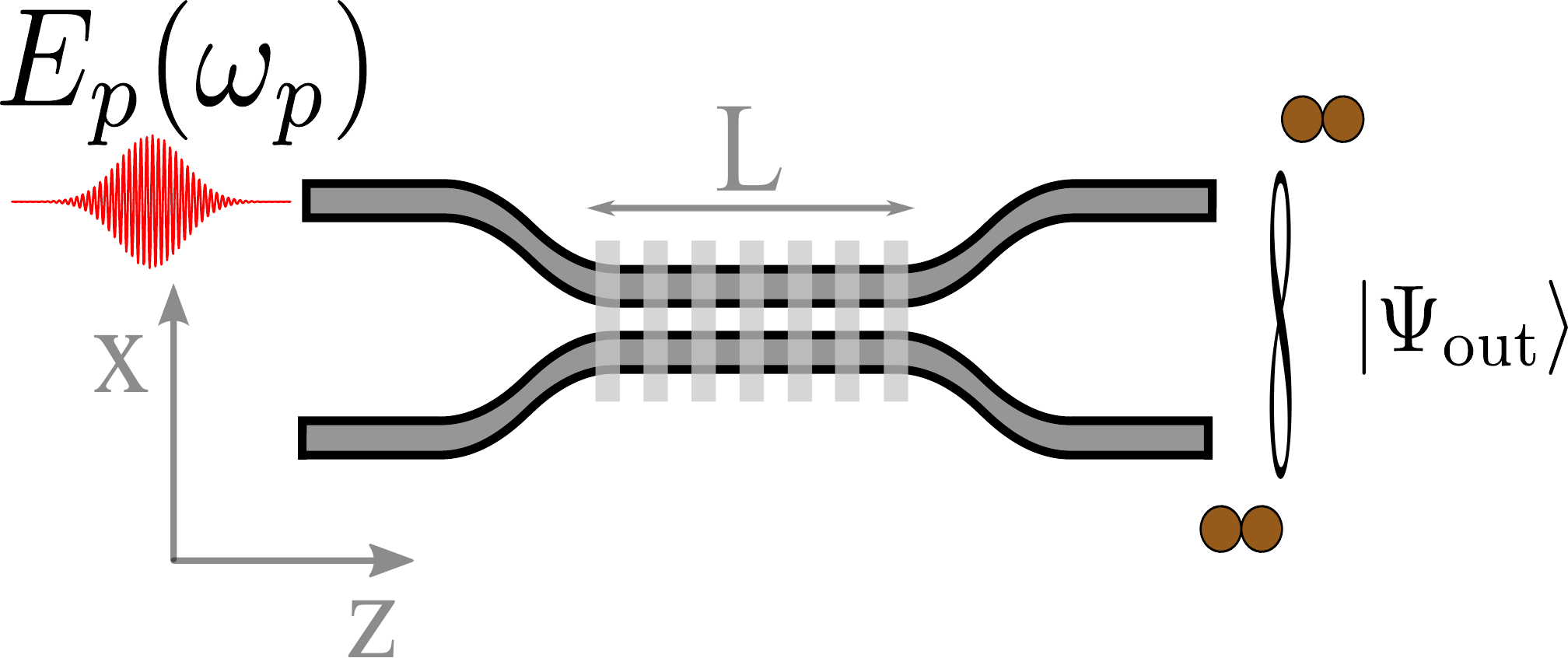}
\caption{Our source consists of a waveguide coupler, where the coupling region of length $L$ is periodically poled. We couple ultrafast pump light into one waveguide of the fan-in region, where it propagates along the $z$-axis. After the periodically poled area, we use the fan-out region to separate the two output ports for our spatially entangled 2-photon N00N state $\ket {\Psi_\mathrm{out}}$.}
\label{fig:sys_sketch}
\end{figure}

The system which we are considering is sketched in figure \ref{fig:sys_sketch}.
The underlying waveguide structure is that of a directional coupler. It consists of a fan-in region for easier access to the waveguide inputs and a coupling region in the middle, where the two waveguides run parallely at a short distance of a few $\mu$m. The strength of the coupling, described by the coupling parameter $C$ is directly given by the distance between the two waveguides and the operating wavelength $\lambda$ of the directional coupler. The coupling region is then followed by a fan-out region to separate the two waveguide outputs again. We define that the fields propagate along the $z$-axis, while coupling happens in $x$-direction.

So far this optical element is purely linear in nature. However, we add a periodic poling to the coupling part of length $L$ in the directional coupler. This allows for parametric down-conversion (PDC) \cite{louisell_quantum_1961, burnham_observation_1970} to take place for a specific parameter combination. We assume, that only the generated quantum fields in the telecom regime are affected by the directional coupler, while the pump stays unmodified in the pumped waveguide channel.

\subsection{Linear Analysis}

First, we need to review the linear properties of a directional coupler, as they are the key to understanding the unique properties of the non-linear process in this structure.

From integrated optics it is long known, how to describe the linear properties of a directional coupler \cite{somekh_channel_1973, marom_relation_1984}. Here, we use the coupled-mode approximation, where the description of the coupled system is given via a linear combination of the modes in the uncoupled system. Solving the linear differential equation of the electric fields in the coupled waveguide system yields two eigenmodes with two non-degenerate eigenvalues
\begin{equation}
\begin{aligned}
i\beta_{A} &= i[\beta^{(0)}+C]\\
i\beta_{S} &= i[\beta^{(0)}-C]\, ,
\end{aligned}
\label{eq:eigenvalues}
\end{equation}
where $\beta^{(0)}$ is the propagation constant of the uncoupled waveguide.
We choose the labels $A$ and $S$ based on the shape (antisymmetric, symmetric) of the eigenmodes in the coupler structure. They are in this approximation given by a linear combination of the spatial modes of the electric field $E$ in the uncoupled waveguides 1 and 2
\begin{equation}
\begin{aligned}
E_A &= \frac{1}{\sqrt{2}} (E_1-E_2)\\
E_S &= \frac{1}{\sqrt{2}} (E_1+E_2)\, .
\end{aligned}
\label{eq:eigenmodes}
\end{equation}
This result is not only a solution to the coupled differential equation for the coupler system, but it also gives instruction on how to switch between the waveguide basis, which is our measurement basis, and the eigenmode basis of the directional coupler.

\subsection{Non-Linear Analysis}

Using the eigenmode description we calculate the generated PDC state in a directional coupler following the approach of \cite{christ_spatial_2009, solntsev_spontaneous_2012, kruse_spatio-spectral_2013}. To compute the generated bi-photon state in the directional coupler, we express the electric fields in the PDC-Hamiltonian
\begin{equation}
\h{H}_{PDC} = \chi^{(2)}\int_V d^3r (\mathcal{E}_p^{(+)} \h{E}^{(-)}\h{E}^{(-)} + h.c.) 
\end{equation}
with the eigenmodes of the coupler, as they diagonalise the linear part of the full system Hamiltonian. Here, $\chi^{(2)}$ denotes the strength of the second order non-linear coefficient, $\mathcal{E}_p$ the classical pump field and $\h{E}$ the generated quantum fields. As we will only consider type-I PDC, where the photons are fundamentally indistinguishable \cite{loudon_quantum_2000}, the two generated quantum fields can be described by the same operator.
\begin{figure}
\includegraphics[width=.3\textwidth]{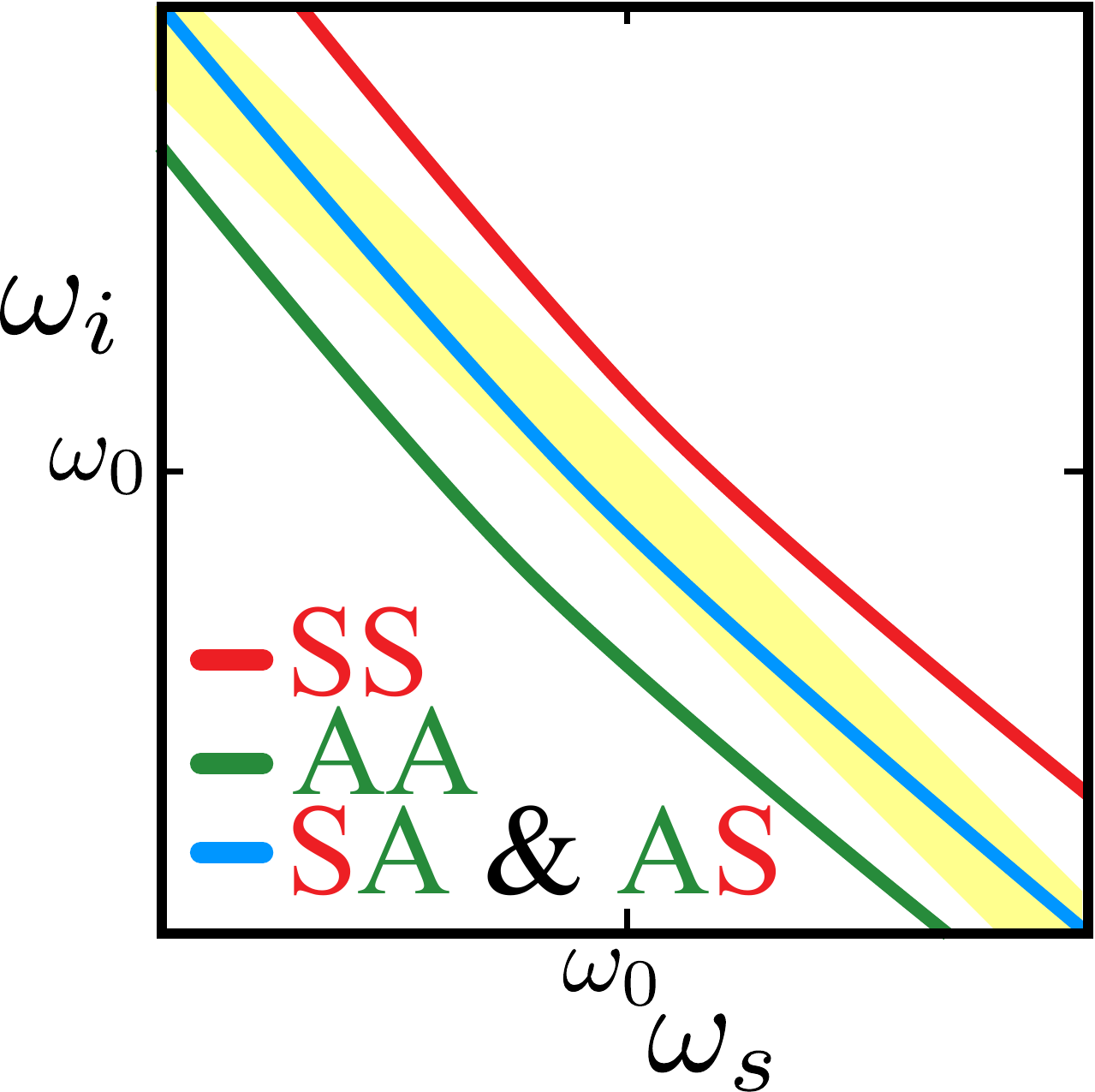}
\caption{The different eigenmode combinations for the photon-pair generation yield different phase-matching conditions, as their propagation constants are modified uniquely. The process generating two photons into the symmetric mode is modified by $-2C$, while two photons into the antisymmetric mode is modified by $+2C$. In the case that one photon is generated into the symmetric and antisymmetric mode (as highlighted by the yellow shading), the modifications cancel each other. $\omega_0$ denotes the degenerate phase-matching condition for signal and idler fields in an uncoupled waveguide.}
\label{fig:phasematching}
\end{figure}
After a lengthy, but straightforward calculation, we arrive at the final PDC state in the eigenmode basis of the waveguide coupler
\begin{equation}
\begin{aligned}
\ket{\Psi}^{\mathrm{Eig.}} &= \frac{1}{\mathcal{N}} \int d\omega_s \int d\omega_i \alpha(\omega_s+\omega_i)\\
&\left[\gamma\, \sinc\left(\Delta\beta_{S,S}\frac{L}{2}\right) e^{-i\Delta\beta_{S,S}\frac{L}{2}} \h{a}_S^\dagger(\omega_s)\h{a}^\dagger_S(\omega_i)\right.\\
&+\delta\, \sinc\left(\Delta\beta_{S,A}\frac{L}{2}\right) e^{-i\Delta\beta_{S,A}\frac{L}{2}} \h{a}_S^\dagger(\omega_s)\h{a}_A^\dagger(\omega_i)\\
&+ \delta\, \sinc\left(\Delta\beta_{A,S}\frac{L}{2}\right) e^{-i\Delta\beta_{A,S}\frac{L}{2}} \h{a}_A^\dagger(\omega_s)\h{a}_S^\dagger(\omega_i)\\
&+\left. \gamma\, \sinc\left(\Delta\beta_{A,A}\frac{L}{2}\right) e^{-i\Delta\beta_{A,A}\frac{L}{2}} \h{a}_A^\dagger(\omega_s)\h{a}_A^\dagger(\omega_i) \right] \ket{0}\, ,
\end{aligned}
\end{equation}
where $\gamma$ and $\delta$ are the pump excitation amplitudes for the symmetric and the antisymmetric eigenmodes, $\alpha(\omega_s+\omega_i)$ is the spectral pump shape depending on the signal and idler photon frequencies $\omega_s$ and $\omega_i$, $\Delta\beta_{ij}$ is the phase-mismatch for the eigenmode combination $(i,j)$,  $\h{a}_k^\dagger(\omega)$ is the creation operator for a photon of frequency $\omega$ in mode $k$ and $\mathcal{N}$ the normalisation constant.
In the coupled system, the PDC process generates the photons into superpositions of the eigenmodes \cite{kruse_spatio-spectral_2013}. As the coupler only features two eigenmodes, there are four combinations generate photons in different eigenmodes. It is only possible to generate either two photons into the symmetric, two photons into the antisymmetric or one photon in each of the two eigenmodes. So far, this is simple combination. However, remembering the linear description of the directional coupler, we find, that these different combinations of eigenmodes correspond to different phase-matching conditions, as the propagation constant is modified uniquely for different eigenmodes. This enables us to specifically excite spatial properties of the generated photon pairs via spectral selection of the corresponding phase-matching condition. In the following, we specifically select the yellow shaded phase-matching condition (corresponding to a selection of the pump wavelength in this region) of figure \ref{fig:phasematching}, which belongs to the generation of 1 photon in each eigenmode. 

While the PDC state in the eigenmode basis of the directional coupler is a full description of the system, we still need to translate the properties of this state into the measurement basis of the laboratory. For this purpose, we choose the waveguide basis and switch back from the eigenmodes using equation \eqref{eq:eigenmodes}. This yields for the yellow shaded phase-matching condition (i.e. only $\Delta \beta_{A,S}\: \&\: \Delta\beta_{S,A}$ contribute to the final state) 
\begin{widetext}
\begin{equation}
\begin{aligned}
\ket{\Psi}^{WG}&=\frac{\delta}{2\sqrt{\mathcal{N}}}\int d\omega_s d\omega_i \alpha(\omega_s+\omega_i)\\
&\left[ \left\{\sinc\left(\Delta\beta_{S,A}\frac{L}{2}\right) e^{-i\Delta\beta_{S,A}\frac{L}{2}}+ \sinc\left(\Delta\beta_{A,S}\frac{L}{2}\right) e^{-i\Delta\beta_{A,S}\frac{L}{2}}
\right\}\h{a}^\dagger_1(\omega_s) \h{a}^\dagger_1(\omega_i)\right. \\
&- \underbrace{\left\{ \sinc\left(\Delta\beta_{S,A}\frac{L}{2}\right) e^{-i\Delta\beta_{S,A}\frac{L}{2}}- \sinc\left(\Delta\beta_{A,S}\frac{L}{2}\right) e^{-i\Delta\beta_{A,S}\frac{L}{2}}
\right\}}_{=0}\h{a}^\dagger_1(\omega_s) \h{a}^\dagger_2(\omega_i) \\
&+\underbrace{\left\{ \sinc\left(\Delta\beta_{S,A}\frac{L}{2}\right) e^{-i\Delta\beta_{S,A}\frac{L}{2}}- \sinc\left(\Delta\beta_{A,S}\frac{L}{2}\right\} e^{-i\Delta\beta_{A,S}\frac{L}{2}}
 \right)}_{=0}\h{a}^\dagger_2(\omega_s) \h{a}^\dagger_1(\omega_i) \\
&-\left\{\left.\sinc\left(\Delta\beta_{S,A}\frac{L}{2}\right) e^{-i\Delta\beta_{S,A}\frac{L}{2}}+ \sinc\left(\Delta\beta_{A,S}\frac{L}{2}\right) e^{-i\Delta\beta_{A,S}\frac{L}{2}}\right\}\h{a}^\dagger_2(\omega_s) \h{a}^\dagger_2(\omega_i) \right] \ket{0} \, ,
\end{aligned}
\end{equation}
\end{widetext}
where $\h{a}^\dagger_k(\omega)$ creates a photon of frequency $\omega$ in waveguide $k$. The key to the N00N state generation is embedded in the two middle terms of this state. We have already stated, that it is possible to simultaneously generate one photon (signal) in the symmetric and the other (idler) in the antisymmetric eigenmode. However, the interchanged combination (idler in symmetric, signal in antisymmetric) is also possible, but with a phase-flip. As these two possibilities are indistinguishable (type-I PDC), the two terms cancel out during the basis transformation. This is the main reason for the post-processing free generation of 2-photon N00N states in this device. To give more physical understanding to this state, we explicitly illustrate the generation principle in the following.

\begin{figure}
\includegraphics[width=.45\textwidth]{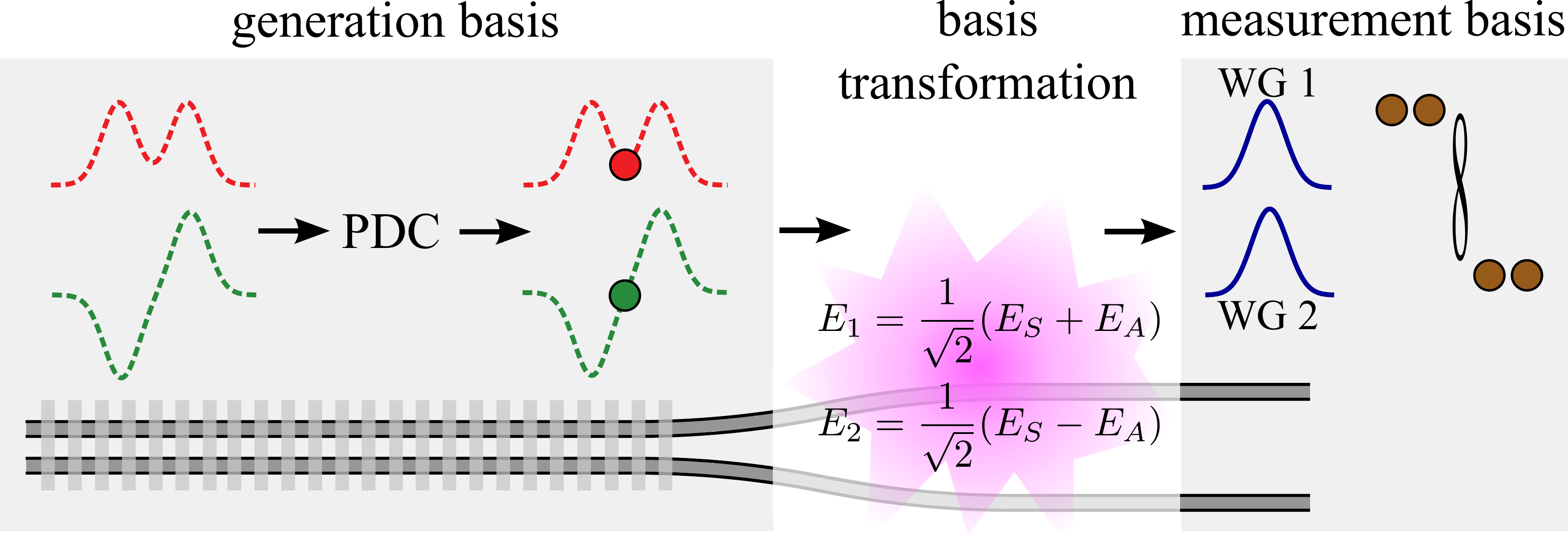}
\caption{ At the beginning of the periodically poled coupler structure, the eigenmodes are unoccupied. During the PDC process, we generate photons with a certain probability only in the combination of symmetric and antisymmetric eigenmode. During the basis transformation a Hong-Ou-Mandel type interference takes place and cancels the coincidences between the two waveguide outputs, leading to the creation of post-processing free 2-photon N00N states.}
\label{fig:generation_protocol}
\end{figure}

Figure \ref{fig:generation_protocol} shows the schematic generation process of photon pairs in the non-linear coupler. 
At the beginning of our waveguide based source, we find the unpopulated eigenmodes of the system, as we are not using a seed to stimulate the process. 
During the PDC process, we find, that we are populating the eigenmodes with one photon each from a generated pair. The probability of this generation is given via the strength of the non-linear interaction.
After the poled region, we have to perform the basis transformation to the measurement basis in our lab. Incidentally, this basis transformation is mathematically fully equivalent to a 50/50 beam splitter with the eigenmodes as inputs and the waveguide modes as outputs. Note, however, that there is no \textit{physical} beam splitter implemented on the chip. Here, only the basis transformation to the measurement basis gives rise to the beam splitter transformation working on our quantum state. Seeing, that we put one photon into each port of our basis beam splitter, we receive Hong-Ou-Mandel type interference \cite{hong_measurement_1987}, cancelling the coincidence contributions between the two waveguide outputs. This results in the post-processing free 2-photon N00N state.

\section{Numerical Results}

In this section we use the analytically calculated output state for the non-linear coupler PDC to calculate the expected fidelity for the 2-photon N00N state, generated by our device.

\begin{figure}
\includegraphics[width=.45\textwidth]{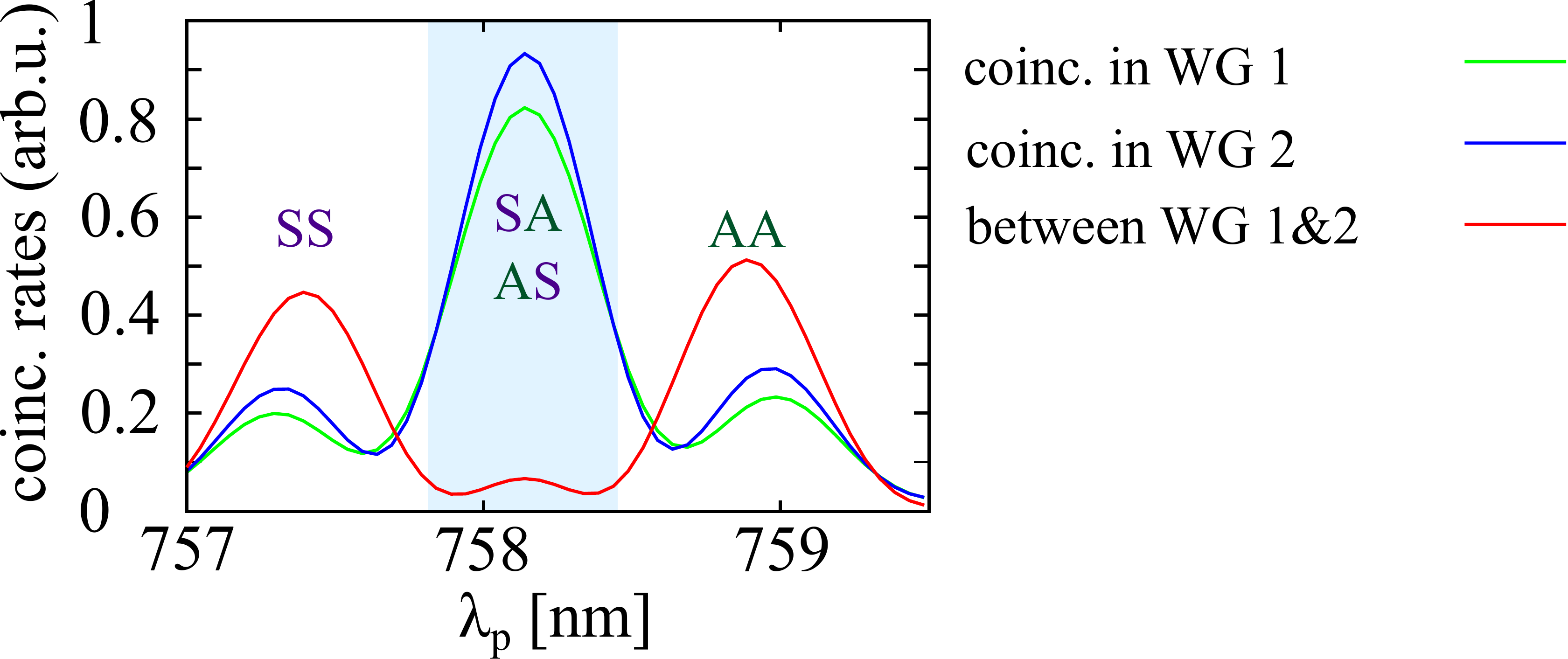}
\caption{The phase-matching scan by tuning the pump wavelength shows a clear suppression of coincidences between the two waveguide ports (red) in the middle of the phase-matching, while coincidences in the single waveguides peak (green: pumped waveguide, blue: unpumped waveguide).}
\label{fig:coincidences}
\end{figure}

For a perfect N00N state, we would expect 2-photon events (coincidences) in either the pumped waveguide (WG 1) or the unpumped one (WG 2) with perfect suppression of coincidences between the waveguides. 
However, we need a careful selection of the pump wavelength to achieve this state, as shown in figure \ref{fig:coincidences}. We tune the pump wavelength to scan the different phase-matching conditions and plot the coincidence rates for the different measurement combinations; coincidences in the pumped (unpumped) waveguide in green (blue) or between waveguides (red). The eigenmode combinations (S, S) and (A, A) show a clear peak for the coincidences between waveguides, while coincidences in a single waveguide show a less pronounced coincidence peak. Using the beam splitter interpretation of the basis transformation this behaviour can be explained easily. For these two eigenmode combinations, we are inserting two photons on the same side of our basis beam splitter and therefore generate coincidences between waveguides with a 50$\,$\% probability. In the middle of the phase-matching roughly at $758\,$nm, exciting the (SA, AS) phase-matching condition, we observe a suppression of coincidence counts between waveguides, a clear sign for the generation of 2-photon N00N states. 
To characterise the fidelity of the generated state, we use the coincidence probabilities for the above mentioned measurement combinations
\begin{equation}
\mathcal{F}=\frac{p_{\mathrm{Coinc,WG 1}}+p_{\mathrm{Coinc,WG 2}}-p_{\mathrm{Coinc,WG 1\&2}}}{p_{\mathrm{Coinc,WG 1}}+p_{\mathrm{Coinc,WG 2}}+p_{\mathrm{Coinc,WG 1\&2}}}\, .
\end{equation}
For the used parameter combination we achieve a fidelity of $\mathcal{F}\approx 93\, \%$. It is restricted by the sinc-sidepeak contributions from the other phase-matching conditions. However, by careful parameter design we can eliminate this restriction on the state fidelity, by e.g. choosing a longer coupler stem length leading to a narrowing the peaks or a higher coupling parameter leading to a larger spectral separation.

\section{Conclusion}
In this paper, we have introduced a single device producing post-processing free 2-photon N00N states. We have discussed the system in detail and have described the generation process of the PDC state. In future, an experimental implementation of this device is planned, showing high state fidelity, as well as the 2-photon N00N state double fringe frequency.

\vspace{0.5cm}

\begin{acknowledgments}
R. K., L. S. and C. S. acknowledge financial support from DFG TRR 142. C. S. H. and I. J. received financial support from Grants No. RVO 68407700 and No. GACR 13-33906 S.
\end{acknowledgments}
\vspace{0.2cm}
\noindent Correspondence should be addressed to:\\ regina.kruse@upb.de

\bibliography{2-wg-arxiv.bib}

\begin{thebibliography}{10}%
\makeatletter
\providecommand \@ifxundefined [1]{%
 \ifx #1\undefined \expandafter \@firstoftwo
 \else \expandafter \@secondoftwo
\fi
}%
\providecommand \@ifnum [1]{%
 \ifnum #1\expandafter \@firstoftwo
 \else \expandafter \@secondoftwo
\fi
}%
\providecommand \enquote [1]{``#1''}%
\providecommand \bibnamefont  [1]{#1}%
\providecommand \bibfnamefont [1]{#1}%
\providecommand \citenamefont [1]{#1}%
\providecommand\href[0]{\@sanitize\@href}%
\providecommand\@href[1]{\endgroup\@@startlink{#1}\endgroup\@@href}%
\providecommand\@@href[1]{#1\@@endlink}%
\providecommand \@sanitize [0]{\begingroup\catcode`\&12\catcode`\#12\relax}%
\@ifxundefined \pdfoutput {\@firstoftwo}{%
 \@ifnum{\z@=\pdfoutput}{\@firstoftwo}{\@secondoftwo}%
}{%
 \providecommand\@@startlink[1]{\leavevmode\special{html:<a href="#1">}}%
 \providecommand\@@endlink[0]{\special{html:</a>}}%
}{%
 \providecommand\@@startlink[1]{%
  \leavevmode
  \pdfstartlink
   attr{/Border[0 0 1 ]/H/I/C[0 1 1]}%
   user{/Subtype/Link/A<</Type/Action/S/URI/URI(#1)>>}%
  \relax
 }%
 \providecommand\@@endlink[0]{\pdfendlink}%
}%
\providecommand \url  [0]{\begingroup\@sanitize \@url }%
\providecommand \@url [1]{\endgroup\@href {#1}{\urlprefix}}%
\providecommand \urlprefix [0]{URL }%
\providecommand \Eprint[0]{\href }%
\@ifxundefined \urlstyle {%
  \providecommand \doi [1]{doi:\discretionary{}{}{}#1}%
}{%
  \providecommand \doi [0]{doi:\discretionary{}{}{}\begingroup
  \urlstyle{rm}\Url }%
}%
\providecommand \doibase [0]{http://dx.doi.org/}%
\providecommand \Doi[1]{\href{\doibase#1}}%
\providecommand \bibAnnote [3]{%
  \BibitemShut{#1}%
  \begin{quotation}\noindent
    \textsc{Key:}\ #2\\\textsc{Annotation:}\ #3%
  \end{quotation}%
}%
\providecommand \bibAnnoteFile [2]{%
  \IfFileExists{#2}{\bibAnnote {#1} {#2} {\input{#2}}}{}%
}%
\providecommand \typeout [0]{\immediate \write \m@ne }%
\providecommand \selectlanguage [0]{\@gobble}%
\providecommand \bibinfo [0]{\@secondoftwo}%
\providecommand \bibfield [0]{\@secondoftwo}%
\providecommand \translation [1]{[#1]}%
\providecommand \BibitemOpen[0]{}%
\providecommand \bibitemStop [0]{}%
\providecommand \bibitemNoStop [0]{.\EOS\space}%
\providecommand \EOS [0]{\spacefactor3000\relax}%
\providecommand \BibitemShut [1]{\csname bibitem#1\endcsname}%
\bibitem{tanzilli_highly_2001}%
  \BibitemOpen
  \bibfield{author}{%
  \bibinfo {author} {\bibfnamefont{S.}~\bibnamefont{Tanzilli}}, \bibinfo
  {author} {\bibfnamefont{H.}~\bibnamefont{De~Riedmatten}}, \bibinfo {author}
  {\bibfnamefont{H.}~\bibnamefont{Tittel}}, \bibinfo {author}
  {\bibfnamefont{H.}~\bibnamefont{Zbinden}}, \bibinfo {author}
  {\bibfnamefont{P.}~\bibnamefont{Baldi}}, \bibinfo {author}
  {\bibfnamefont{M.}~\bibnamefont{De~Micheli}}, \bibinfo {author}
  {\bibfnamefont{D.}~\bibnamefont{Ostrowsky}},\ and\ \bibinfo {author}
  {\bibfnamefont{N.}~\bibnamefont{Gisin}},\ }%
  \bibfield{journal}{%
  \Doi{10.1049/el:20010009}{\bibinfo {journal} {Electronics Letters}}\ }%
  \textbf{\bibinfo {volume} {37}},\ \bibinfo {pages} {26 } (\bibinfo {month}
  {Jan.}\ \bibinfo {year} {2001}),\ ISSN \bibinfo {issn} {0013-5194}%
  \bibAnnoteFile{NoStop}{tanzilli_highly_2001}%
\bibitem{politi_silica--silicon_2008}%
  \BibitemOpen
  \bibfield{author}{%
  \bibinfo {author} {\bibfnamefont{A.}~\bibnamefont{Politi}}, \bibinfo {author}
  {\bibfnamefont{M.~J.}\ \bibnamefont{Cryan}}, \bibinfo {author}
  {\bibfnamefont{J.~G.}\ \bibnamefont{Rarity}}, \bibinfo {author}
  {\bibfnamefont{S.}~\bibnamefont{Yu}},\ and\ \bibinfo {author}
  {\bibfnamefont{J.~L.}\ \bibnamefont{O'Brien}},\ }%
  \bibfield{journal}{%
  \Doi{10.1126/science.1155441}{\bibinfo {journal} {Science}}\ }%
  \textbf{\bibinfo {volume} {320}},\ \bibinfo {pages} {646} (\bibinfo {month}
  {Feb.}\ \bibinfo {year} {2008}),\ ISSN \bibinfo {issn} {0036-8075,
  1095-9203},\ \url{http://www.sciencemag.org/content/320/5876/646}%
  \bibAnnoteFile{NoStop}{politi_silica--silicon_2008}%
\bibitem{peruzzo_quantum_2010}%
  \BibitemOpen
  \bibfield{author}{%
  \bibinfo {author} {\bibfnamefont{A.}~\bibnamefont{Peruzzo}}, \bibinfo
  {author} {\bibfnamefont{M.}~\bibnamefont{Lobino}}, \bibinfo {author}
  {\bibfnamefont{J.~C.~F.}\ \bibnamefont{Matthews}}, \bibinfo {author}
  {\bibfnamefont{N.}~\bibnamefont{Matsuda}}, \bibinfo {author}
  {\bibfnamefont{A.}~\bibnamefont{Politi}}, \bibinfo {author}
  {\bibfnamefont{K.}~\bibnamefont{Poulios}}, \bibinfo {author}
  {\bibfnamefont{X.-Q.}\ \bibnamefont{Zhou}}, \bibinfo {author}
  {\bibfnamefont{Y.}~\bibnamefont{Lahini}}, \bibinfo {author}
  {\bibfnamefont{N.}~\bibnamefont{Ismail}}, \bibinfo {author}
  {\bibfnamefont{K.}~\bibnamefont{Wörhoff}}, \bibinfo {author}
  {\bibfnamefont{Y.}~\bibnamefont{Bromberg}}, \bibinfo {author}
  {\bibfnamefont{Y.}~\bibnamefont{Silberberg}}, \bibinfo {author}
  {\bibfnamefont{M.~G.}\ \bibnamefont{Thompson}},\ and\ \bibinfo {author}
  {\bibfnamefont{J.~L.}\ \bibnamefont{OBrien}},\ }%
  \bibfield{journal}{%
  \Doi{10.1126/science.1193515}{\bibinfo {journal} {Science}}\ }%
  \textbf{\bibinfo {volume} {329}},\ \bibinfo {pages} {1500} (\bibinfo {month}
  {Sep.}\ \bibinfo {year} {2010}),\ ISSN \bibinfo {issn} {0036-8075,
  1095-9203},\ \url{http://www.sciencemag.org/content/329/5998/1500}%
  \bibAnnoteFile{NoStop}{peruzzo_quantum_2010}%
\bibitem{owens_two-photon_2011}%
  \BibitemOpen
  \bibfield{author}{%
  \bibinfo {author} {\bibfnamefont{J.~O.}\ \bibnamefont{Owens}}, \bibinfo
  {author} {\bibfnamefont{M.~A.}\ \bibnamefont{Broome}}, \bibinfo {author}
  {\bibfnamefont{D.~N.}\ \bibnamefont{Biggerstaff}}, \bibinfo {author}
  {\bibfnamefont{M.~E.}\ \bibnamefont{Goggin}}, \bibinfo {author}
  {\bibfnamefont{A.}~\bibnamefont{Fedrizzi}}, \bibinfo {author}
  {\bibfnamefont{T.}~\bibnamefont{Linjordet}}, \bibinfo {author}
  {\bibfnamefont{M.}~\bibnamefont{Ams}}, \bibinfo {author}
  {\bibfnamefont{G.~D.}\ \bibnamefont{Marshall}}, \bibinfo {author}
  {\bibfnamefont{J.}~\bibnamefont{Twamley}}, \bibinfo {author}
  {\bibfnamefont{M.~J.}\ \bibnamefont{Withford}},\ and\ \bibinfo {author}
  {\bibfnamefont{A.~G.}\ \bibnamefont{White}},\ }%
  \bibfield{journal}{%
  \Doi{10.1088/1367-2630/13/7/075003}{\bibinfo {journal} {New Journal of
  Physics}}\ }%
  \textbf{\bibinfo {volume} {13}},\ \bibinfo {pages} {075003} (\bibinfo {month}
  {Jul.}\ \bibinfo {year} {2011}),\ ISSN \bibinfo {issn} {1367-2630},\
  \url{http://iopscience.iop.org/1367-2630/13/7/075003}%
  \bibAnnoteFile{NoStop}{owens_two-photon_2011}%
\bibitem{crespi_integrated_2011}%
  \BibitemOpen
  \bibfield{author}{%
  \bibinfo {author} {\bibfnamefont{A.}~\bibnamefont{Crespi}}, \bibinfo {author}
  {\bibfnamefont{R.}~\bibnamefont{Ramponi}}, \bibinfo {author}
  {\bibfnamefont{R.}~\bibnamefont{Osellame}}, \bibinfo {author}
  {\bibfnamefont{L.}~\bibnamefont{Sansoni}}, \bibinfo {author}
  {\bibfnamefont{I.}~\bibnamefont{Bongioanni}}, \bibinfo {author}
  {\bibfnamefont{F.}~\bibnamefont{Sciarrino}}, \bibinfo {author}
  {\bibfnamefont{G.}~\bibnamefont{Vallone}},\ and\ \bibinfo {author}
  {\bibfnamefont{P.}~\bibnamefont{Mataloni}},\ }%
  \bibfield{journal}{%
  \Doi{10.1038/ncomms1570}{\bibinfo {journal} {Nature Communications}}\ }%
  \textbf{\bibinfo {volume} {2}},\ \bibinfo {pages} {566} (\bibinfo {month}
  {Nov.}\ \bibinfo {year} {2011}),\
  \url{http://www.nature.com/ncomms/journal/v2/n11/full/ncomms1570.html}%
  \bibAnnoteFile{NoStop}{crespi_integrated_2011}%
\bibitem{krapick_efficient_2013}%
  \BibitemOpen
  \bibfield{author}{%
  \bibinfo {author} {\bibfnamefont{S.}~\bibnamefont{Krapick}}, \bibinfo
  {author} {\bibfnamefont{H.}~\bibnamefont{Herrmann}}, \bibinfo {author}
  {\bibfnamefont{V.}~\bibnamefont{Quiring}}, \bibinfo {author}
  {\bibfnamefont{B.}~\bibnamefont{Brecht}}, \bibinfo {author}
  {\bibfnamefont{H.}~\bibnamefont{Suche}},\ and\ \bibinfo {author}
  {\bibfnamefont{C.}~\bibnamefont{Silberhorn}},\ }%
  \bibfield{journal}{%
  \Doi{10.1088/1367-2630/15/3/033010}{\bibinfo {journal} {New Journal of
  Physics}}\ }%
  \textbf{\bibinfo {volume} {15}},\ \bibinfo {pages} {033010} (\bibinfo {month}
  {Mar.}\ \bibinfo {year} {2013}),\ ISSN \bibinfo {issn} {1367-2630},\
  \url{http://iopscience.iop.org/1367-2630/15/3/033010}%
  \bibAnnoteFile{NoStop}{krapick_efficient_2013}%
\bibitem{corrielli_rotated_2014}%
  \BibitemOpen
  \bibfield{author}{%
  \bibinfo {author} {\bibfnamefont{G.}~\bibnamefont{Corrielli}}, \bibinfo
  {author} {\bibfnamefont{A.}~\bibnamefont{Crespi}}, \bibinfo {author}
  {\bibfnamefont{R.}~\bibnamefont{Geremia}}, \bibinfo {author}
  {\bibfnamefont{R.}~\bibnamefont{Ramponi}}, \bibinfo {author}
  {\bibfnamefont{L.}~\bibnamefont{Sansoni}}, \bibinfo {author}
  {\bibfnamefont{A.}~\bibnamefont{Santinelli}}, \bibinfo {author}
  {\bibfnamefont{P.}~\bibnamefont{Mataloni}}, \bibinfo {author}
  {\bibfnamefont{F.}~\bibnamefont{Sciarrino}},\ and\ \bibinfo {author}
  {\bibfnamefont{R.}~\bibnamefont{Osellame}},\ }%
  \bibfield{journal}{%
  \bibinfo {journal} {Nature Communications}\ }%
  \textbf{\bibinfo {volume} {5}} (\bibinfo {month} {Jun.}\ \bibinfo {year}
  {2014}),\ \doi{\bibinfo {doi} {10.1038/ncomms5249}},\
  \url{http://www.nature.com/ncomms/2014/140625/ncomms5249/full/ncomms5249.htm%
l}%
  \bibAnnoteFile{NoStop}{corrielli_rotated_2014}%
\bibitem{metcalf_quantum_2014}%
  \BibitemOpen
  \bibfield{author}{%
  \bibinfo {author} {\bibfnamefont{B.~J.}\ \bibnamefont{Metcalf}}, \bibinfo
  {author} {\bibfnamefont{J.~B.}\ \bibnamefont{Spring}}, \bibinfo {author}
  {\bibfnamefont{P.~C.}\ \bibnamefont{Humphreys}}, \bibinfo {author}
  {\bibfnamefont{N.}~\bibnamefont{Thomas-Peter}}, \bibinfo {author}
  {\bibfnamefont{M.}~\bibnamefont{Barbieri}}, \bibinfo {author}
  {\bibfnamefont{W.~S.}\ \bibnamefont{Kolthammer}}, \bibinfo {author}
  {\bibfnamefont{X.-M.}\ \bibnamefont{Jin}}, \bibinfo {author}
  {\bibfnamefont{N.~K.}\ \bibnamefont{Langford}}, \bibinfo {author}
  {\bibfnamefont{D.}~\bibnamefont{Kundys}}, \bibinfo {author}
  {\bibfnamefont{J.~C.}\ \bibnamefont{Gates}}, \bibinfo {author}
  {\bibfnamefont{B.~J.}\ \bibnamefont{Smith}}, \bibinfo {author}
  {\bibfnamefont{P.~G.~R.}\ \bibnamefont{Smith}},\ and\ \bibinfo {author}
  {\bibfnamefont{I.~A.}\ \bibnamefont{Walmsley}},\ }%
  \bibfield{journal}{%
  \Doi{10.1038/nphoton.2014.217}{\bibinfo {journal} {Nature Photonics}}\ }%
  \textbf{\bibinfo {volume} {8}},\ \bibinfo {pages} {770} (\bibinfo {month}
  {Oct.}\ \bibinfo {year} {2014}),\ ISSN \bibinfo {issn} {1749-4885},\
  \url{http://www.nature.com/nphoton/journal/v8/n10/full/nphoton.2014.217.html%
}%
  \bibAnnoteFile{NoStop}{metcalf_quantum_2014}%
\bibitem{silverstone_-chip_2014}%
  \BibitemOpen
  \bibfield{author}{%
  \bibinfo {author} {\bibfnamefont{J.~W.}\ \bibnamefont{Silverstone}}, \bibinfo
  {author} {\bibfnamefont{D.}~\bibnamefont{Bonneau}}, \bibinfo {author}
  {\bibfnamefont{K.}~\bibnamefont{Ohira}}, \bibinfo {author}
  {\bibfnamefont{N.}~\bibnamefont{Suzuki}}, \bibinfo {author}
  {\bibfnamefont{H.}~\bibnamefont{Yoshida}}, \bibinfo {author}
  {\bibfnamefont{N.}~\bibnamefont{Iizuka}}, \bibinfo {author}
  {\bibfnamefont{M.}~\bibnamefont{Ezaki}}, \bibinfo {author}
  {\bibfnamefont{C.~M.}\ \bibnamefont{Natarajan}}, \bibinfo {author}
  {\bibfnamefont{M.~G.}\ \bibnamefont{Tanner}}, \bibinfo {author}
  {\bibfnamefont{R.~H.}\ \bibnamefont{Hadfield}}, \bibinfo {author}
  {\bibfnamefont{V.}~\bibnamefont{Zwiller}}, \bibinfo {author}
  {\bibfnamefont{G.~D.}\ \bibnamefont{Marshall}}, \bibinfo {author}
  {\bibfnamefont{J.~G.}\ \bibnamefont{Rarity}}, \bibinfo {author}
  {\bibfnamefont{J.~L.}\ \bibnamefont{O'Brien}},\ and\ \bibinfo {author}
  {\bibfnamefont{M.~G.}\ \bibnamefont{Thompson}},\ }%
  \bibfield{journal}{%
  \Doi{10.1038/nphoton.2013.339}{\bibinfo {journal} {Nature Photonics}}\ }%
  \textbf{\bibinfo {volume} {8}},\ \bibinfo {pages} {104} (\bibinfo {month}
  {Feb.}\ \bibinfo {year} {2014}),\ ISSN \bibinfo {issn} {1749-4885},\
  \url{http://www.nature.com/nphoton/journal/v8/n2/full/nphoton.2013.339.html}%
  \bibAnnoteFile{NoStop}{silverstone_-chip_2014}%
\bibitem{jin_-chip_2014}%
  \BibitemOpen
  \bibfield{author}{%
  \bibinfo {author} {\bibfnamefont{H.}~\bibnamefont{Jin}}, \bibinfo {author}
  {\bibfnamefont{F.}~\bibnamefont{Liu}}, \bibinfo {author}
  {\bibfnamefont{P.}~\bibnamefont{Xu}}, \bibinfo {author}
  {\bibfnamefont{J.}~\bibnamefont{Xia}}, \bibinfo {author}
  {\bibfnamefont{M.}~\bibnamefont{Zhong}}, \bibinfo {author}
  {\bibfnamefont{Y.}~\bibnamefont{Yuan}}, \bibinfo {author}
  {\bibfnamefont{J.}~\bibnamefont{Zhou}}, \bibinfo {author}
  {\bibfnamefont{Y.}~\bibnamefont{Gong}}, \bibinfo {author}
  {\bibfnamefont{W.}~\bibnamefont{Wang}},\ and\ \bibinfo {author}
  {\bibfnamefont{S.}~\bibnamefont{Zhu}},\ }%
  \bibfield{journal}{%
  \Doi{10.1103/PhysRevLett.113.103601}{\bibinfo {journal} {Physical Review
  Letters}}\ }%
  \textbf{\bibinfo {volume} {113}},\ \bibinfo {pages} {103601} (\bibinfo
  {month} {Sep.}\ \bibinfo {year} {2014}),\
  \url{http://link.aps.org/doi/10.1103/PhysRevLett.113.103601}%
  \bibAnnoteFile{NoStop}{jin_-chip_2014}%
\bibitem{crespi_integrated_2013}%
  \BibitemOpen
  \bibfield{author}{%
  \bibinfo {author} {\bibfnamefont{A.}~\bibnamefont{Crespi}}, \bibinfo {author}
  {\bibfnamefont{R.}~\bibnamefont{Osellame}}, \bibinfo {author}
  {\bibfnamefont{R.}~\bibnamefont{Ramponi}}, \bibinfo {author}
  {\bibfnamefont{D.~J.}\ \bibnamefont{Brod}}, \bibinfo {author}
  {\bibfnamefont{E.~F.}\ \bibnamefont{Galvão}}, \bibinfo {author}
  {\bibfnamefont{N.}~\bibnamefont{Spagnolo}}, \bibinfo {author}
  {\bibfnamefont{C.}~\bibnamefont{Vitelli}}, \bibinfo {author}
  {\bibfnamefont{E.}~\bibnamefont{Maiorino}}, \bibinfo {author}
  {\bibfnamefont{P.}~\bibnamefont{Mataloni}},\ and\ \bibinfo {author}
  {\bibfnamefont{F.}~\bibnamefont{Sciarrino}},\ }%
  \bibfield{journal}{%
  \Doi{10.1038/nphoton.2013.112}{\bibinfo {journal} {Nature Photonics}}\ }%
  \textbf{\bibinfo {volume} {7}},\ \bibinfo {pages} {545} (\bibinfo {month}
  {Jul.}\ \bibinfo {year} {2013}),\ ISSN \bibinfo {issn} {1749-4885},\
  \url{http://www.nature.com/nphoton/journal/v7/n7/abs/nphoton.2013.112.html}%
  \bibAnnoteFile{NoStop}{crespi_integrated_2013}%
\bibitem{spring_boson_2013}%
  \BibitemOpen
  \bibfield{author}{%
  \bibinfo {author} {\bibfnamefont{J.~B.}\ \bibnamefont{Spring}}, \bibinfo
  {author} {\bibfnamefont{B.~J.}\ \bibnamefont{Metcalf}}, \bibinfo {author}
  {\bibfnamefont{P.~C.}\ \bibnamefont{Humphreys}}, \bibinfo {author}
  {\bibfnamefont{W.~S.}\ \bibnamefont{Kolthammer}}, \bibinfo {author}
  {\bibfnamefont{X.-M.}\ \bibnamefont{Jin}}, \bibinfo {author}
  {\bibfnamefont{M.}~\bibnamefont{Barbieri}}, \bibinfo {author}
  {\bibfnamefont{A.}~\bibnamefont{Datta}}, \bibinfo {author}
  {\bibfnamefont{N.}~\bibnamefont{Thomas-Peter}}, \bibinfo {author}
  {\bibfnamefont{N.~K.}\ \bibnamefont{Langford}}, \bibinfo {author}
  {\bibfnamefont{D.}~\bibnamefont{Kundys}}, \bibinfo {author}
  {\bibfnamefont{J.~C.}\ \bibnamefont{Gates}}, \bibinfo {author}
  {\bibfnamefont{B.~J.}\ \bibnamefont{Smith}}, \bibinfo {author}
  {\bibfnamefont{P.~G.~R.}\ \bibnamefont{Smith}},\ and\ \bibinfo {author}
  {\bibfnamefont{I.~A.}\ \bibnamefont{Walmsley}},\ }%
  \bibfield{journal}{%
  \Doi{10.1126/science.1231692}{\bibinfo {journal} {Science}}\ }%
  \textbf{\bibinfo {volume} {339}},\ \bibinfo {pages} {798} (\bibinfo {month}
  {Feb.}\ \bibinfo {year} {2013}),\ ISSN \bibinfo {issn} {0036-8075,
  1095-9203},\ \url{http://www.sciencemag.org/content/339/6121/798}%
  \bibAnnoteFile{NoStop}{spring_boson_2013}%
\bibitem{broome_photonic_2013}%
  \BibitemOpen
  \bibfield{author}{%
  \bibinfo {author} {\bibfnamefont{M.~A.}\ \bibnamefont{Broome}}, \bibinfo
  {author} {\bibfnamefont{A.}~\bibnamefont{Fedrizzi}}, \bibinfo {author}
  {\bibfnamefont{S.}~\bibnamefont{Rahimi-Keshari}}, \bibinfo {author}
  {\bibfnamefont{J.}~\bibnamefont{Dove}}, \bibinfo {author}
  {\bibfnamefont{S.}~\bibnamefont{Aaronson}}, \bibinfo {author}
  {\bibfnamefont{T.~C.}\ \bibnamefont{Ralph}},\ and\ \bibinfo {author}
  {\bibfnamefont{A.~G.}\ \bibnamefont{White}},\ }%
  \bibfield{journal}{%
  \Doi{10.1126/science.1231440}{\bibinfo {journal} {Science}}\ }%
  \textbf{\bibinfo {volume} {339}},\ \bibinfo {pages} {794} (\bibinfo {month}
  {Feb.}\ \bibinfo {year} {2013}),\ ISSN \bibinfo {issn} {0036-8075,
  1095-9203},\ \url{http://www.sciencemag.org/content/339/6121/794}%
  \bibAnnoteFile{NoStop}{broome_photonic_2013}%
\bibitem{tillmann_experimental_2013}%
  \BibitemOpen
  \bibfield{author}{%
  \bibinfo {author} {\bibfnamefont{M.}~\bibnamefont{Tillmann}}, \bibinfo
  {author} {\bibfnamefont{B.}~\bibnamefont{Dakić}}, \bibinfo {author}
  {\bibfnamefont{R.}~\bibnamefont{Heilmann}}, \bibinfo {author}
  {\bibfnamefont{S.}~\bibnamefont{Nolte}}, \bibinfo {author}
  {\bibfnamefont{A.}~\bibnamefont{Szameit}},\ and\ \bibinfo {author}
  {\bibfnamefont{P.}~\bibnamefont{Walther}},\ }%
  \bibfield{journal}{%
  \bibinfo {journal} {Nature Photonics}\ }%
  \textbf{\bibinfo {volume} {advance online publication}} (\bibinfo {month}
  {May}\ \bibinfo {year} {2013}),\ ISSN \bibinfo {issn} {1749-4893},\
  \doi{\bibinfo {doi} {10.1038/nphoton.2013.102}},\
  \url{http://www.nature.com/nphoton/journal/vaop/ncurrent/full/nphoton.2013.1%
02.html}%
  \bibAnnoteFile{NoStop}{tillmann_experimental_2013}%
\bibitem{mista_nonclassical_1997}%
  \BibitemOpen
  \bibfield{author}{%
  \bibinfo {author} {\bibfnamefont{L.}~\bibnamefont{Mišta}}\ and\ \bibinfo
  {author} {\bibfnamefont{J.}~\bibnamefont{Peřina}},\ }%
  \bibfield{journal}{%
  \Doi{10.1023/A:1021116819329}{\bibinfo {journal} {Czechoslovak Journal of
  Physics}}\ }%
  \textbf{\bibinfo {volume} {47}},\ \bibinfo {pages} {629} (\bibinfo {month}
  {Jun.}\ \bibinfo {year} {1997}),\ ISSN \bibinfo {issn} {0011-4626,
  1572-9486},\
  \url{http://link.springer.com/article/10.1023/A%3A1021116819329}%
  \bibAnnoteFile{NoStop}{mista_nonclassical_1997}%
\bibitem{solntsev_spontaneous_2012}%
  \BibitemOpen
  \bibfield{author}{%
  \bibinfo {author} {\bibfnamefont{A.~S.}\ \bibnamefont{Solntsev}}, \bibinfo
  {author} {\bibfnamefont{A.~A.}\ \bibnamefont{Sukhorukov}}, \bibinfo {author}
  {\bibfnamefont{D.~N.}\ \bibnamefont{Neshev}},\ and\ \bibinfo {author}
  {\bibfnamefont{Y.~S.}\ \bibnamefont{Kivshar}},\ }%
  \bibfield{journal}{%
  \Doi{10.1103/PhysRevLett.108.023601}{\bibinfo {journal} {Physical Review
  Letters}}\ }%
  \textbf{\bibinfo {volume} {108}},\ \bibinfo {pages} {023601} (\bibinfo
  {month} {Jan.}\ \bibinfo {year} {2012}),\
  \url{http://link.aps.org/doi/10.1103/PhysRevLett.108.023601}%
  \bibAnnoteFile{NoStop}{solntsev_spontaneous_2012}%
\bibitem{kruse_spatio-spectral_2013}%
  \BibitemOpen
  \bibfield{author}{%
  \bibinfo {author} {\bibfnamefont{R.}~\bibnamefont{Kruse}}, \bibinfo {author}
  {\bibfnamefont{F.}~\bibnamefont{Katzschmann}}, \bibinfo {author}
  {\bibfnamefont{A.}~\bibnamefont{Christ}}, \bibinfo {author}
  {\bibfnamefont{A.}~\bibnamefont{Schreiber}}, \bibinfo {author}
  {\bibfnamefont{S.}~\bibnamefont{Wilhelm}}, \bibinfo {author}
  {\bibfnamefont{K.}~\bibnamefont{Laiho}}, \bibinfo {author}
  {\bibfnamefont{A.}~\bibnamefont{Gábris}}, \bibinfo {author}
  {\bibfnamefont{C.~S.}\ \bibnamefont{Hamilton}}, \bibinfo {author}
  {\bibfnamefont{I.}~\bibnamefont{Jex}},\ and\ \bibinfo {author}
  {\bibfnamefont{C.}~\bibnamefont{Silberhorn}},\ }%
  \bibfield{journal}{%
  \Doi{10.1088/1367-2630/15/8/083046}{\bibinfo {journal} {New Journal of
  Physics}}\ }%
  \textbf{\bibinfo {volume} {15}},\ \bibinfo {pages} {083046} (\bibinfo {month}
  {Aug.}\ \bibinfo {year} {2013}),\ ISSN \bibinfo {issn} {1367-2630},\
  \url{http://iopscience.iop.org/1367-2630/15/8/083046}%
  \bibAnnoteFile{NoStop}{kruse_spatio-spectral_2013}%
\bibitem{solntsev_generation_2014}%
  \BibitemOpen
  \bibfield{author}{%
  \bibinfo {author} {\bibfnamefont{A.~S.}\ \bibnamefont{Solntsev}}, \bibinfo
  {author} {\bibfnamefont{F.}~\bibnamefont{Setzpfandt}}, \bibinfo {author}
  {\bibfnamefont{A.~S.}\ \bibnamefont{Clark}}, \bibinfo {author}
  {\bibfnamefont{C.~W.}\ \bibnamefont{Wu}}, \bibinfo {author}
  {\bibfnamefont{M.~J.}\ \bibnamefont{Collins}}, \bibinfo {author}
  {\bibfnamefont{C.}~\bibnamefont{Xiong}}, \bibinfo {author}
  {\bibfnamefont{A.}~\bibnamefont{Schreiber}}, \bibinfo {author}
  {\bibfnamefont{F.}~\bibnamefont{Katzschmann}}, \bibinfo {author}
  {\bibfnamefont{F.}~\bibnamefont{Eilenberger}}, \bibinfo {author}
  {\bibfnamefont{R.}~\bibnamefont{Schiek}}, \bibinfo {author}
  {\bibfnamefont{W.}~\bibnamefont{Sohler}}, \bibinfo {author}
  {\bibfnamefont{A.}~\bibnamefont{Mitchell}}, \bibinfo {author}
  {\bibfnamefont{C.}~\bibnamefont{Silberhorn}}, \bibinfo {author}
  {\bibfnamefont{B.~J.}\ \bibnamefont{Eggleton}}, \bibinfo {author}
  {\bibfnamefont{T.}~\bibnamefont{Pertsch}}, \bibinfo {author}
  {\bibfnamefont{A.~A.}\ \bibnamefont{Sukhorukov}}, \bibinfo {author}
  {\bibfnamefont{D.~N.}\ \bibnamefont{Neshev}},\ and\ \bibinfo {author}
  {\bibfnamefont{Y.~S.}\ \bibnamefont{Kivshar}},\ }%
  \bibfield{journal}{%
  \Doi{10.1103/PhysRevX.4.031007}{\bibinfo {journal} {Physical Review X}}\ }%
  \textbf{\bibinfo {volume} {4}},\ \bibinfo {pages} {031007} (\bibinfo {month}
  {Jul.}\ \bibinfo {year} {2014}),\
  \url{http://link.aps.org/doi/10.1103/PhysRevX.4.031007}%
  \bibAnnoteFile{NoStop}{solntsev_generation_2014}%
\bibitem{hamilton_driven_2014}%
  \BibitemOpen
  \bibfield{author}{%
  \bibinfo {author} {\bibfnamefont{C.~S.}\ \bibnamefont{Hamilton}}, \bibinfo
  {author} {\bibfnamefont{R.}~\bibnamefont{Kruse}}, \bibinfo {author}
  {\bibfnamefont{L.}~\bibnamefont{Sansoni}}, \bibinfo {author}
  {\bibfnamefont{C.}~\bibnamefont{Silberhorn}},\ and\ \bibinfo {author}
  {\bibfnamefont{I.}~\bibnamefont{Jex}},\ }%
  \bibfield{journal}{%
  \Doi{10.1103/PhysRevLett.113.083602}{\bibinfo {journal} {Physical Review
  Letters}}\ }%
  \textbf{\bibinfo {volume} {113}},\ \bibinfo {pages} {083602} (\bibinfo
  {month} {Aug.}\ \bibinfo {year} {2014}),\
  \url{http://link.aps.org/doi/10.1103/PhysRevLett.113.083602}%
  \bibAnnoteFile{NoStop}{hamilton_driven_2014}%
\bibitem{lugani_generation_2011}%
  \BibitemOpen
  \bibfield{author}{%
  \bibinfo {author} {\bibfnamefont{J.}~\bibnamefont{Lugani}}, \bibinfo {author}
  {\bibfnamefont{S.}~\bibnamefont{Ghosh}},\ and\ \bibinfo {author}
  {\bibfnamefont{K.}~\bibnamefont{Thyagarajan}},\ }%
  \bibfield{journal}{%
  \Doi{10.1103/PhysRevA.83.062333}{\bibinfo {journal} {Physical Review A}}\ }%
  \textbf{\bibinfo {volume} {83}},\ \bibinfo {pages} {062333} (\bibinfo {month}
  {Jun.}\ \bibinfo {year} {2011}),\
  \url{http://link.aps.org/doi/10.1103/PhysRevA.83.062333}%
  \bibAnnoteFile{NoStop}{lugani_generation_2011}%
\bibitem{dowling_quantum_2008}%
  \BibitemOpen
  \bibfield{author}{%
  \bibinfo {author} {\bibfnamefont{J.~P.}\ \bibnamefont{Dowling}},\ }%
  \bibfield{journal}{%
  \Doi{10.1080/00107510802091298}{\bibinfo {journal} {Contemporary Physics}}\
  }%
  \textbf{\bibinfo {volume} {49}},\ \bibinfo {pages} {125} (\bibinfo {month}
  {Mar.}\ \bibinfo {year} {2008}),\ ISSN \bibinfo {issn} {0010-7514,
  1366-5812},\
  \url{http://www.tandfonline.com/doi/abs/10.1080/.VARtL9aP-PI#.VAcCCK2P-PI}%
  \bibAnnoteFile{NoStop}{dowling_quantum_2008}%
\bibitem{afek_high-noon_2010}%
  \BibitemOpen
  \bibfield{author}{%
  \bibinfo {author} {\bibfnamefont{I.}~\bibnamefont{Afek}}, \bibinfo {author}
  {\bibfnamefont{O.}~\bibnamefont{Ambar}},\ and\ \bibinfo {author}
  {\bibfnamefont{Y.}~\bibnamefont{Silberberg}},\ }%
  \bibfield{journal}{%
  \Doi{10.1126/science.1188172}{\bibinfo {journal} {Science}}\ }%
  \textbf{\bibinfo {volume} {328}},\ \bibinfo {pages} {879} (\bibinfo {month}
  {May}\ \bibinfo {year} {2010}),\ ISSN \bibinfo {issn} {0036-8075,
  1095-9203},\ \url{http://www.sciencemag.org/content/328/5980/879}%
  \bibAnnoteFile{NoStop}{afek_high-noon_2010}%
\bibitem{israel_experimental_2012}%
  \BibitemOpen
  \bibfield{author}{%
  \bibinfo {author} {\bibfnamefont{Y.}~\bibnamefont{Israel}}, \bibinfo {author}
  {\bibfnamefont{I.}~\bibnamefont{Afek}}, \bibinfo {author}
  {\bibfnamefont{S.}~\bibnamefont{Rosen}}, \bibinfo {author}
  {\bibfnamefont{O.}~\bibnamefont{Ambar}},\ and\ \bibinfo {author}
  {\bibfnamefont{Y.}~\bibnamefont{Silberberg}},\ }%
  \bibfield{journal}{%
  \Doi{10.1103/PhysRevA.85.022115}{\bibinfo {journal} {Physical Review A}}\ }%
  \textbf{\bibinfo {volume} {85}},\ \bibinfo {pages} {022115} (\bibinfo {month}
  {Feb.}\ \bibinfo {year} {2012}),\
  \url{http://link.aps.org/doi/10.1103/PhysRevA.85.022115}%
  \bibAnnoteFile{NoStop}{israel_experimental_2012}%
\bibitem{boyd_quantum_2012}%
  \BibitemOpen
  \bibfield{author}{%
  \bibinfo {author} {\bibfnamefont{R.~W.}\ \bibnamefont{Boyd}}\ and\ \bibinfo
  {author} {\bibfnamefont{J.~P.}\ \bibnamefont{Dowling}},\ }%
  \bibfield{journal}{%
  \Doi{10.1007/s11128-011-0253-y}{\bibinfo {journal} {Quantum Information
  Processing}}\ }%
  \textbf{\bibinfo {volume} {11}},\ \bibinfo {pages} {891} (\bibinfo {month}
  {Aug.}\ \bibinfo {year} {2012}),\ ISSN \bibinfo {issn} {1570-0755,
  1573-1332},\
  \url{http://link.springer.com/article/10.1007/s11128-011-0253-y}%
  \bibAnnoteFile{NoStop}{boyd_quantum_2012}%
\bibitem{louisell_quantum_1961}%
  \BibitemOpen
  \bibfield{author}{%
  \bibinfo {author} {\bibfnamefont{W.~H.}\ \bibnamefont{Louisell}}, \bibinfo
  {author} {\bibfnamefont{A.}~\bibnamefont{Yariv}},\ and\ \bibinfo {author}
  {\bibfnamefont{A.~E.}\ \bibnamefont{Siegman}},\ }%
  \bibfield{journal}{%
  \Doi{10.1103/PhysRev.124.1646}{\bibinfo {journal} {Physical Review}}\ }%
  \textbf{\bibinfo {volume} {124}},\ \bibinfo {pages} {1646} (\bibinfo {month}
  {Dec.}\ \bibinfo {year} {1961}),\
  \url{http://link.aps.org/doi/10.1103/PhysRev.124.1646}%
  \bibAnnoteFile{NoStop}{louisell_quantum_1961}%
\bibitem{burnham_observation_1970}%
  \BibitemOpen
  \bibfield{author}{%
  \bibinfo {author} {\bibfnamefont{D.~C.}\ \bibnamefont{Burnham}}\ and\
  \bibinfo {author} {\bibfnamefont{D.~L.}\ \bibnamefont{Weinberg}},\ }%
  \bibfield{journal}{%
  \Doi{10.1103/PhysRevLett.25.84}{\bibinfo {journal} {Physical Review
  Letters}}\ }%
  \textbf{\bibinfo {volume} {25}},\ \bibinfo {pages} {84} (\bibinfo {month}
  {Jul.}\ \bibinfo {year} {1970}),\
  \url{http://link.aps.org/doi/10.1103/PhysRevLett.25.84}%
  \bibAnnoteFile{NoStop}{burnham_observation_1970}%
\bibitem{somekh_channel_1973}%
  \BibitemOpen
  \bibfield{author}{%
  \bibinfo {author} {\bibfnamefont{S.}~\bibnamefont{Somekh}}, \bibinfo {author}
  {\bibfnamefont{E.}~\bibnamefont{Garmire}}, \bibinfo {author}
  {\bibfnamefont{A.}~\bibnamefont{Yariv}}, \bibinfo {author}
  {\bibfnamefont{H.~L.}\ \bibnamefont{Garvin}},\ and\ \bibinfo {author}
  {\bibfnamefont{R.~G.}\ \bibnamefont{Hunsperger}},\ }%
  \bibfield{journal}{%
  \Doi{10.1063/1.1654468}{\bibinfo {journal} {Applied Physics Letters}}\ }%
  \textbf{\bibinfo {volume} {22}},\ \bibinfo {pages} {46} (\bibinfo {month}
  {Jan.}\ \bibinfo {year} {1973}),\ ISSN \bibinfo {issn} {0003-6951,
  1077-3118},\
  \url{http://scitation.aip.org/content/aip/journal/apl/22/1/10.1063/1.1654468%
}%
  \bibAnnoteFile{NoStop}{somekh_channel_1973}%
\bibitem{marom_relation_1984}%
  \BibitemOpen
  \bibfield{author}{%
  \bibinfo {author} {\bibfnamefont{E.}~\bibnamefont{Marom}}, \bibinfo {author}
  {\bibfnamefont{O.}~\bibnamefont{Ramer}},\ and\ \bibinfo {author}
  {\bibfnamefont{S.}~\bibnamefont{Ruschin}},\ }%
  \bibfield{journal}{%
  \Doi{10.1109/JQE.1984.1072326}{\bibinfo {journal} {{IEEE} Journal of Quantum
  Electronics}}\ }%
  \textbf{\bibinfo {volume} {20}},\ \bibinfo {pages} {1311} (\bibinfo {month}
  {Dec.}\ \bibinfo {year} {1984}),\ ISSN \bibinfo {issn} {0018-9197}%
  \bibAnnoteFile{NoStop}{marom_relation_1984}%
\bibitem{christ_spatial_2009}%
  \BibitemOpen
  \bibfield{author}{%
  \bibinfo {author} {\bibfnamefont{A.}~\bibnamefont{Christ}}, \bibinfo {author}
  {\bibfnamefont{K.}~\bibnamefont{Laiho}}, \bibinfo {author}
  {\bibfnamefont{A.}~\bibnamefont{Eckstein}}, \bibinfo {author}
  {\bibfnamefont{T.}~\bibnamefont{Lauckner}}, \bibinfo {author}
  {\bibfnamefont{P.~J.}\ \bibnamefont{Mosley}},\ and\ \bibinfo {author}
  {\bibfnamefont{C.}~\bibnamefont{Silberhorn}},\ }%
  \bibfield{journal}{%
  \bibinfo {journal} {Phys. Rev. A}\ }%
  \textbf{\bibinfo {volume} {80}},\ \bibinfo {pages} {033829} (\bibinfo {year}
  {2009})%
  \bibAnnoteFile{NoStop}{christ_spatial_2009}%
\bibitem{loudon_quantum_2000}%
  \BibitemOpen
  \bibfield{author}{%
  \bibinfo {author} {\bibfnamefont{R.}~\bibnamefont{Loudon}},\ }%
  \emph{\bibinfo {title} {The Quantum Theory of Light}}\ (\bibinfo {publisher}
  {Oxford University Press},\ \bibinfo {year} {2000})\ ISBN \bibinfo {isbn}
  {9780198501770}%
  \bibAnnoteFile{NoStop}{loudon_quantum_2000}%
\bibitem{hong_measurement_1987}%
  \BibitemOpen
  \bibfield{author}{%
  \bibinfo {author} {\bibfnamefont{C.~K.}\ \bibnamefont{Hong}}, \bibinfo
  {author} {\bibfnamefont{Z.~Y.}\ \bibnamefont{Ou}},\ and\ \bibinfo {author}
  {\bibfnamefont{L.}~\bibnamefont{Mandel}},\ }%
  \bibfield{journal}{%
  \Doi{10.1103/PhysRevLett.59.2044}{\bibinfo {journal} {Physical Review
  Letters}}\ }%
  \textbf{\bibinfo {volume} {59}},\ \bibinfo {pages} {2044} (\bibinfo {month}
  {Nov.}\ \bibinfo {year} {1987}),\
  \url{http://link.aps.org/doi/10.1103/PhysRevLett.59.2044}%
  \bibAnnoteFile{NoStop}{hong_measurement_1987}%
\end{thebibliography}%

\end{document}